\documentclass[structabstract]{aa}
\newcommand{\Cho}{\textit{Chorizos} }

\usepackage{txfonts}
\usepackage{graphicx}
\usepackage{natbib} 
\bibpunct{(}{)}{;}{a}{}{,}

\begin{document}

%\title{An HST study of the age distribution of the Orion Nebula Cluster}
\title{Quantitative Evidence for an Intrinsic Age Spread in the Orion Nebula Cluster}

\author{M. Reggiani\inst{\ref{inst1}} \and M. Robberto\inst{\ref{inst2}} \and N. Da Rio\inst{\ref{inst2}} \and M. R. Meyer\inst{\ref{inst1}}  \and D. R. Soderblom\inst{\ref{inst2}}  \and L. Ricci\inst{\ref{inst3}}}

\institute{Institute of Astronomy, ETH Zurich, CH-8093 Zurich, Switzerland\label{inst1} \and Space Telescope Science Institute, 3700 San Martin Drive, Baltimore, MD 21218, USA\label{inst2}  \and European Southern Observatory, Karl-Schwarzschild-Strasse 2, 85748 Garching, Germany\label{inst3}}

\abstract {} {We present a study of the distribution of stellar ages in the Orion Nebula Cluster (ONC) based on accurate HST photometry taken from the HST Treasury Program observations of the ONC utilizing the most recent estimate of the cluster's distance (Menten et al. 2007). We investigate the presence of an intrinsic age spread in the region and a possible trend of age with the spatial distribution.} {We estimate the extinction and accretion luminosity towards each source by performing synthetic photometry on an empirical calibration of atmospheric models (Da Rio et al. 2010) using the package \Cho (Maíz-Apellaníz 2004). The position of the sources in the HR-diagram is compared with different theoretical isochrones to estimate the mean cluster age and age dispersion.  Through Monte Carlo simulations we quantify the amount of intrinsic age spread in the region, taking into account uncertainties on the distance, spectral type, extinction, unresolved binaries, accretion and photometric variability.} {According to Siess et al. (2000) evolutionary models the mean age of the Cluster is 2.2 Myr with a scatter of few Myrs. With Monte Carlo simulations we find that the observed age spread is inconsistent with a coeval stellar population, but is in agreement with a star formation activity between 1.5 and 3.5 Myrs. We also observe light evidence for a trend of ages with spatial distribution.} {}

\keywords{Stars: formation - Stars: low-mass - Stars: pre-main sequence - open clusters and associations: individual: Orion Nebula Cluster - Methods: statistical - Techniques: photometric}
\titlerunning{Age Spread in the ONC} 
\authorrunning{M. Reggiani et al.}
\maketitle
% =========================
% 1. INTRODUCTION
% =========================

\section{Introduction}
The analysis of the age distribution of pre-Main Sequence (PMS) stars in young clusters represents the most direct way of deducing how star formation proceeds in molecular clouds \citep{Hartmann2001,Hillenbrand2008,DaRioB2010}. Ages of young stellar objects are generally estimated from the comparison between the position of individual stars on the HR-diagram and theoretical isochrones and evolutionary tracks. For the majority of stars,  the main uncertainties on age are related to the determination of the intrinsic luminosity, since  the pre-main-sequence evolution of low mass stars takes place at nearly constant effective temperature. As pointed out by \citet{Hartmann2001}, the main sources of error on the luminosity are relative to the source distance, photometric variability, extinction, unresolved binary companions, and accretion.
The discrepancies between different families of pre-main-sequence evolutionary models add further uncertainty, in particular the different slopes of the isochrone lines in the HR\ diagrams make it difficult to unequivocally derive not only the absolute but also the relative ages of cluster members.

Moreover, other physical processes may affect the intrinsic luminosity of PMS stars. \cite{Baraffe2009} present calculations where episodic accretion at early stages of evolution of low-mass stars provides a possible explanation for the observed spread in the HR diagram. \cite{Hosokawa2011} show that, even though episodic accretion may occur, it is not able to explain the observed scatter. In fact, the spread obtained varying the initial radius or the thermal efficiency is much larger than the spread that results from different accretion histories, even though it is not enough to account for all the observed spread in the HR diagram \citep{Hosokawa2011}. However, on one hand, a model quantifying whether the accretion-induced luminosity spread is able to account for the observations has not been created yet. On the other, we lack a direct observational proof for this scenario. Therefore we still interpret the distribution of radii for a given mass as a spread in age.

In this paper we revisit this problem on the basis of new data from the Orion Nebula Cluster obtained with the Hubble Space Telescope (HST). The same issue was partially discussed by \cite{DaRio2010}, using ground based photometry. Our present goal is to assess the possibility of inferring robust age distributions by using high precision multi-band photometry. Fitting all
available photometric data at the same time, we can have accurate estimates for extinction and accretion luminosity which allows us to better constrain the errors on the luminosity and the age of each source.
We also focus on the assessment of apparent age spread induced by these errors.
After a brief description of the photometry used and the selection criteria of our source  sample (Sect. \ref{Photometry}), we present the analysis carried out to assemble an observed HR diagram for the ONC (Sect. \ref{HR diagram}). The study of the relative uncertainties and the results obtained are described in Sect. \ref{source_error} and Sect. \ref{distribution}.
Finally, we discuss the possibility of a true age spread in the Cluster (Sect. \ref{age_spread}) and the dependence of age on the spatial distribution (Sect. \ref{spatial_age_distribution}).

% =========================
% 2. PHOTOMETRY
% =========================

\section{Photometry} \label{Photometry}
The data used for the present work consist mainly of broad-band photometry extracted from the large data-set of ACS/WFC\ observations collected for the HST\ Treasury Program on the Orion Nebula Cluster (GO-10246, PI\ M. Robberto). This program has covered an area of about 450 arcmin$^{2}$, roughly centered on the Trapezium Cluster in five filters. In Table \ref{filters} we list the ACS filters, their ground-based telescope equivalents and the duration of individual exposures. The complete ACS/WFC catalog for the ONC includes $\sim$3200 sources (Robberto 2011, in preparation) that have been classified as either a single star, binary star, photo-evaporated disk, dark silhouette, or possible background galaxy.  A subset of the catalog  describing the non stellar objects has been presented by \citet{Ricci2008}.

The deep ACS observations, designed to sample the substellar component of the cluster,  do not provide photometry for the brightest sources because of saturation. We have therefore complemented the ACS\ survey with ancillary ground based data, collected within the same HST\ Treasury Program. These  observations were obtained simultaneously at the MPG-ESO 2.2-m Telescope at La Silla with the Wide Field Imager (WFI) \citep{DaRio2009} and with the Infrared Side Port Imager (ISPI)  at the CTIO Blanco 4-m Telescope \citep{Robberto2010}, fully covering the area imaged with the HST. In particular, for our analysis we selected the WFI\ data, which were taken in the U, B, V and I passbands, in a narrow-band filter centered on the $H_{\alpha}$ line and in a medium resolution filter centered at 6200 \AA.  This provides a R-band line-free equivalent passband for stars K-type and earlier, whereas for stars M-type and later it is centered on a temperature sensitive TiO absorption band which allows a reddening independent measurement of $T_{eff}$ for cool stars \citep{DaRio2009}. WFI measures  were used only for stars not present in the ACS dataset. To avoid introducing color errors due to source variability, we have not combined the ACS\ and WFI data, as they were obtained at different epochs.

In order to reliably build the HR diagram one has to combine the photometric observations with spectral types.
We have therefore down selected from our original sample of 2793 objects 807 stars with spectral type given by \citet{Hillenbrand1997}, 100 stars with spectral type from a follow-up to the study of \cite{Stassun1999} (see also \cite{DaRio2010}), and 150 stars with spectral type from \cite{DaRio2009}, obtained with narrow-band photometry at 6200 $\AA$ on a TiO absorption feature. In total we have therefore 1057 stars with known spectral type and photometry in at least three bands at optical wavelengths. The completeness of our spectroscopic sample will be analyzed in Sect.~\ref{completeness}.

\begin{table} 
\caption{ACS/WFC filters} 
\label{filters}	% is used to refer this table in the text
\centering \begin{tabular}{c c c} \hline\hline
Filter & Ground Equivalent & Exposure time (s)\\
\hline
F435W & Johnson B & 420 \\
F555W & Johnson V & 385 \\
F658N & H$\alpha$ +[N \begin{small}II\end{small}] $\lambda$ 6583 & 340 \\
F775W & Cousin IC & 385 \\
F850LP & z-Band & 385 \\
\hline
\end{tabular} 
\end{table}

% =========================
% 3. HR DIAGRAM
% =========================

\section{HR diagram} \label{HR diagram}
\subsection{Construction}
The construction of an accurate HR diagram requires the knowledge of several quantities and transformation relations. First, the spectral type has to be converted to effective temperature. For this purpose we used the relation between spectral type and effective temperature given by \citet{Luhman1999}, which is tuned to bring the presumably coeval components of the young system GG Tau on the same isochrone, according to  the  \citet{Baraffe1998} models. For M-type PMS\ stars this relation is intermediate between dwarfs and giants, and as it has been shown by \cite{DaRio2009}, it represents an improvement over the relation assumed by \citet{Hillenbrand1997}. Using this transformation the temperatures of our sample range from $\sim 2500$~K to $\sim 8000$~K.
Second, to estimate the bolometric luminosity of each source one has to convert the observed magnitudes to absolute magnitudes adopting values for extinction, distance modulus, and bolometric corrections. One should also account for non-photospheric sources of luminosity like, e.g., accretion luminosity.

For the ACS\ data we used as a reference the F850LP filter ($z$-band equivalent) using the following relation:
\begin{eqnarray}
&\log\left(L_{\star}/L_{\odot}\right)=  0.4 \cdot    \nonumber \\
&\left[z_{ACS,\odot}-(z_{{\tt ACS},\star}-\Delta {z}_{acc}-A_{z}+BC_{z,\star}(T)-DM)\right]\nonumber
\end{eqnarray}
where $z_{{\tt ACS},\odot}$ is the absolute magnitude of the Sun, $z_{{\tt ACS},\star}$ is the magnitude of the target star, $\Delta {z}_{acc}$ is the estimated contribution to the observed brightness due to disk accretion, $A_{z}$ is the extinction, $BC_{z,\star}$ is the bolometric correction and $DM$ the distance modulus. For the WFI\ data we used an analogous at $z$ expression for the $I$ band.

To estimate the various terms we computed synthetic photometry on families of stellar atmospheric models using the SED fitting package \Cho \citep{Maiz2004}. 
The $\chi^{2}$-minimization SED-fitting enables us to asses the correspondence between observed and synthetic photometry up to N=4 physical parameters. In other words, \Cho solves the problem of finding which model SED is the most compatible with the observed colors in a model N-dimensional parameter space.
We assembled spectral libraries from various authors: \citet{Kurucz1993} for effective temperatures larger than $8000$ $K$,  NextGen model atmospheres  \citep{Haushildt1999} for $5000$ $K$ $\leq T\leq$ $8000$ $K$ and AMES-MT 2000 models \citep{Allard2000} for $T\leq 5000$, all of them with solar metallicity,  following \cite{DaRio2010}.

\citet{DaRio2010} show that colors of ONC members differ from those of main-sequence dwarfs in the M-type regime as well as the atmospheric models do not accurately match the photometry of the ONC .
Therefore, we  limited the parameter space to models corresponding to the empirical calibration suggested by \cite{DaRio2010} as a good approximation for the ONC. This empirical isochrone was built to obtain a good match for a selected sample of non-accreting low-extinction members of the ONC, using the ZAMS intrinsic colors for $T\geq 3700$~K, the 2 Myr synthetic colors for $T\leq 2900$~K, and a linear interpolation between the two for $2900 \geq T$ $\leq 3700$~K. 

Moreover to account for the accretion luminosity that may affect our young  sources, we added a typical accretion spectrum to the family of template spectra.
According to \citet{Calvet1998}, the total accretion column emission can be well approximated by the superposition
of an optically thick emission from the heated photosphere below the shock with an optically thin emission
generated in the infalling flow, in a 3:1 ratio. The optically thick component can be modeled by a black body
at $T_{eff} = 6000-8000$~K, and  we assumed $T_{eff}=7000$~K. For the optically thin emission we used  {\sl Cloudy} \citep{Ferland1998} to generate the spectrum of an  optically thin slab with $n = 10^{8} cm^{-3}$ also at $T\simeq7000$~K . 
We added the combined spectrum to our reference stellar models with ratios of
$\log(L_{accr}/L_{tot})$ increasing from -3 to 0. 

\begin{figure}
\includegraphics[scale=0.40]{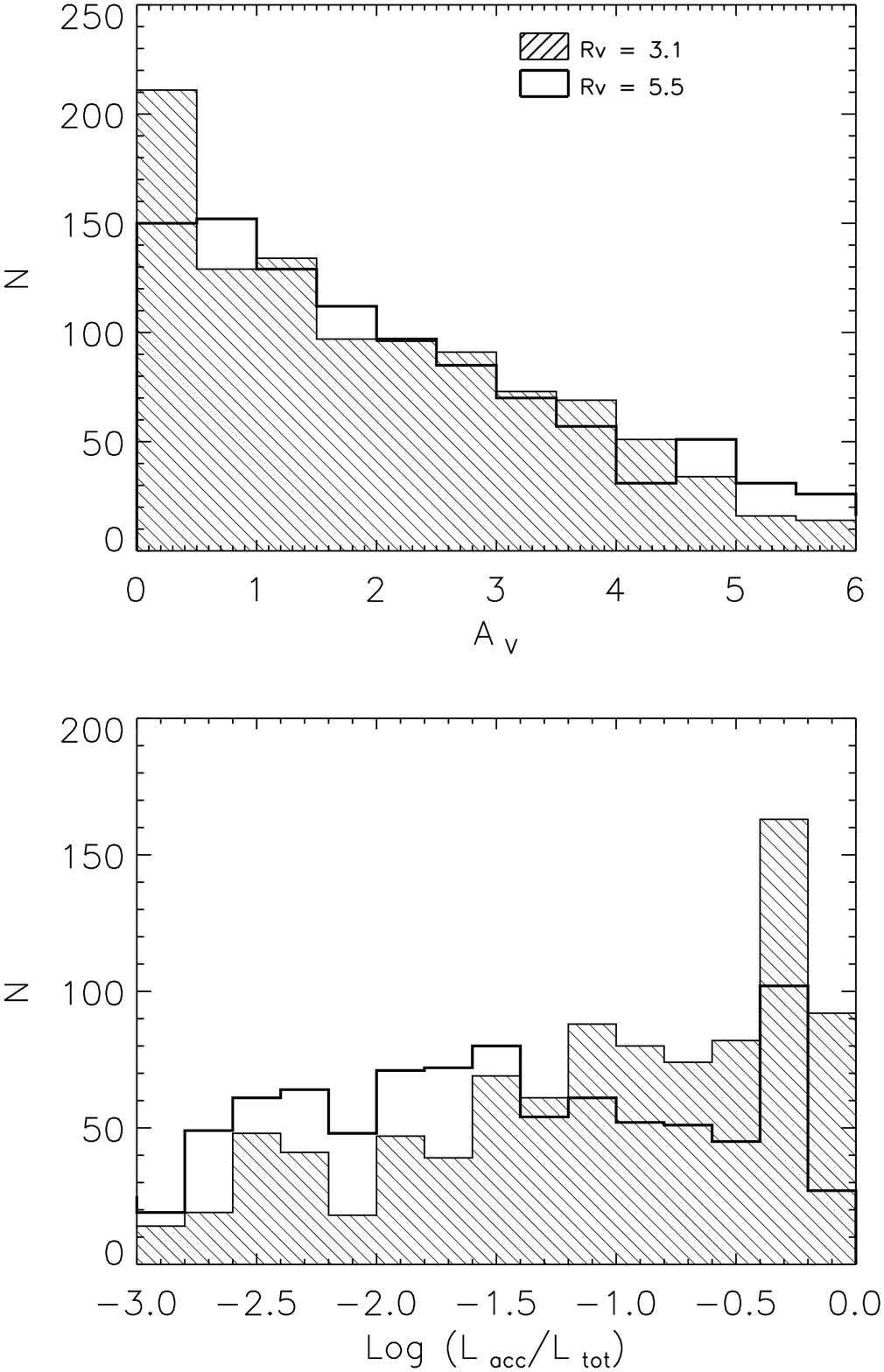}
\caption{\small{\textit{Extinction and Accretion Luminosity}. Top: Distribution of the extinction for the sample of 1057  stars from the ACS and WFI catalogs for which we have estimated extinction and accretion luminosity.  Bottom: Distribution of the accretion luminosity for the same sample of objects. The accretion luminosity is expressed in terms of $L_{accr}/L_{tot}$ and in logarithmic scale ranges from -3 to -0.}}
\label{histo_av}
\end{figure}

We ran \Cho for each source of our sample, assuming the metallicity and the stellar effective temperature as fixed parameters and two different reddening laws, corresponding to the standard $R_{V}=3.1$ and $R_{V}=5.5$, by some considered more appropriate for the Orion Nebula \citep{Johnson1967,Costero1970}. Therefore, extinction and accretion represent the variables of our fitting procedure. \Cho returned the most likely value of extinction and accretion luminosity for each source, for each reddening law. With Chorizos we can fit all available photometric data at the same time to derive extinction and accretion luminosity for each source, whereas \cite{DaRio2010} concentrated on 2 colors.
Chorizos is therefore more robust against the occasional spurious data points, or against the data with the highest photometric error.
In Fig.~\ref{histo_av} we show the distributions of the extinction and accretion luminosity for the two choices of reddening law. 
For the standard reddening law, that we will adopt as main reference, the extinction distribution has a median value at $A_{V}=2.2$ mag and shows that roughly 70\% of the sources are extincted by less than 3 mag.  Concerning the distribution of $(L_{accr}/L_{tot})$, we find an average value for the relative fraction of 5\% and more than half of the stars in the sample showing an excess attributed to accretion with $L_{accr}/L_{tot}$ greater than 3\% (comparable to what was found by \cite{DaRio2010}).

The bolometric corrections appropriate  for our
population were derived directly from the integration of the synthetic spectra in $z$ as well as across the whole range of wavelengths:
\begin{equation}\label{BC}
 BC(T)=-2.5 \log \left[\frac{\int_{\lambda}S_{\lambda}
F_{\lambda}(T)d\lambda}{\int_{\lambda} F_{\lambda}(T)
d\lambda}/\frac{\int_{\lambda} S_{\lambda}
F_{\lambda}(\odot)d\lambda}{\int_{\lambda} F_{\lambda}(\odot)
d\lambda}\right]
\end{equation}
where $F_{\lambda}(T)$ is the spectral energy distribution of the star
of temperature $T$, while $F_{\lambda}(\odot)$ is the solar spectral
profile, and $S_{\lambda}$ is the passband of the filter multiplied by the quantum efficiency function of the detector.
According to Equation~\ref{BC}, the bolometric correction of the Sun is
zero in each band, whereas the zero-points in this case are irrelevant as they cancel out in the calculation of the luminosity.

We also integrated the fluxes of a  NextGen synthetic solar spectrum,  assuming the canonical value of $M_{V,\odot}$ (V Johnson)= +4.83 \citep{Binney1998}.
The absolute magnitudes of the Sun  in the $z$ and $I$ bands of ACS and WFI turn out, respectively, $M_{z}=4.002$ and $M_{I}=4.020$, computed with synthetic photometry on solar spectrum.
Concerning the distance value, we assumed  $d=414\pm7$ pc according to \cite{Menten2007}.
This corresponds to a distance modulus $DM=8.085$.

\subsection{Results} \label{sub_results}
Having estimated the temperature and luminosity for each source, we produced the HR diagrams for the two values of reddening laws mentioned above. We then compared the position of the stars with three sets of evolutionary tracks and isochrones: \citet{Siess2000}, \citet{Palla1999} and \citet{Dantona1997} (extended in 1998).

\begin{figure*}
\resizebox{\hsize}{!}{\includegraphics[scale=0.27]{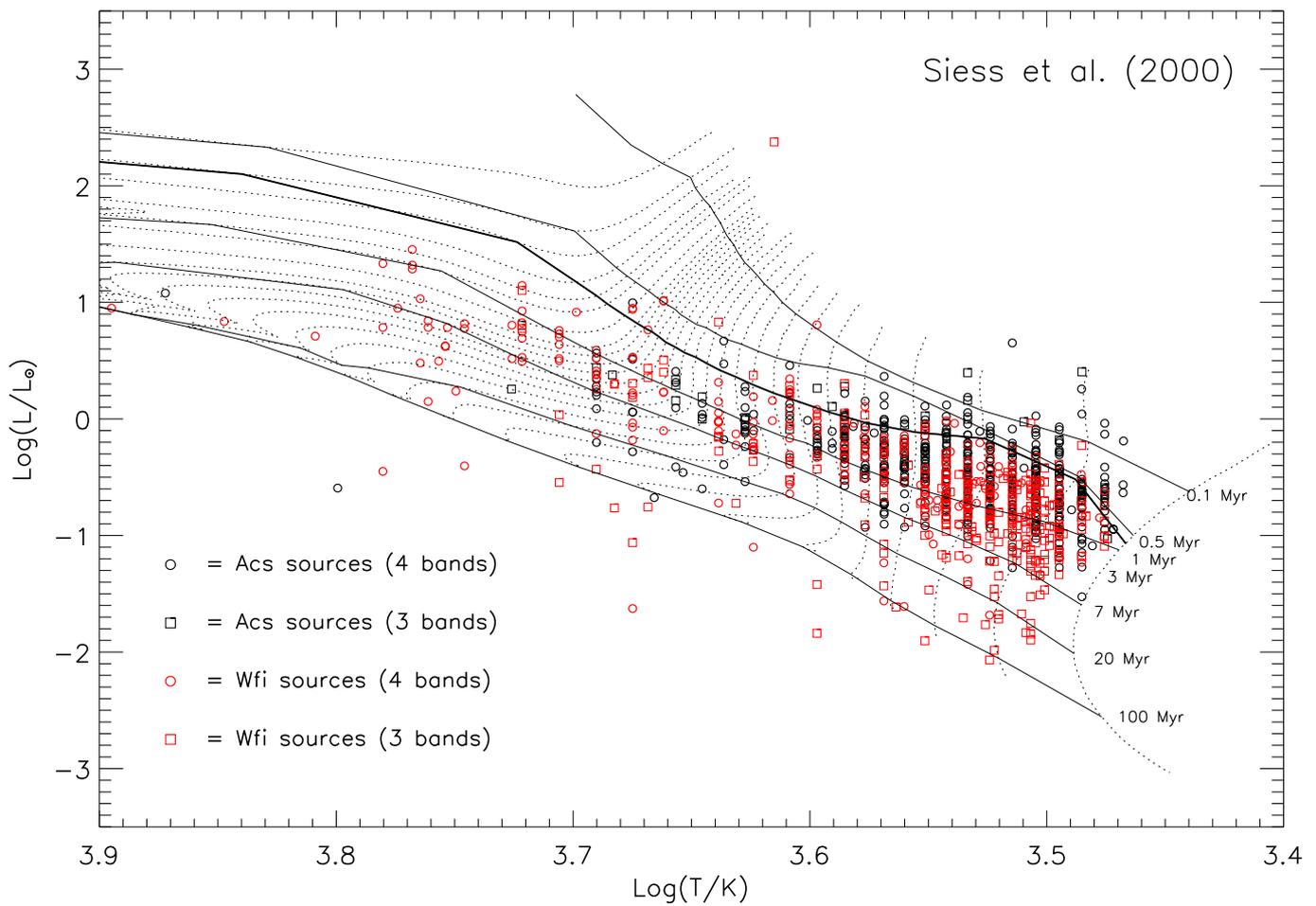}}
\caption{\small{\textit{HR diagrams for the Orion Nebula Cluster}. Black symbols represent ACS sources: circles are the stars with four-bands photometry while squares represent stars with complete three-bands photometry. Red squares indicate respectively WFI objects with 4 bands photometry, while red diamonds the WFI stars with three-bands photometry. Superimposed are the evolutionary calculations by \citet{Siess2000}. Solid lines represent isochrones, the thickest one indicate the 1 Myr isochrone. Dotted lines represent evolutionary tracks and they range from 0.1 up to 5 $M_{\odot}$.}}\label{hr_diagr_siess}
\end{figure*}

In Fig. \ref{hr_diagr_siess} we show the HR diagram relative to the reddening law $R_{V}=3.1$ and \citet{Siess2000} calculations.  Currently there is no family of evolutionary models showing a significantly higher consistency with dynamical mass estimates \citep{Hillenbrand2004}. Therefore we adopt \citet{Siess2000} calculations as our reference, as they provide the smallest apparent age spread on the HR diagram. Moreover, according to \citet{DaRio2010}, these models reduce the systematic correlation of average age with mass, more prominent when using \cite{Palla1999} isochrones, or \citet{Dantona1994} models \citep{Hillenbrand1997}.
In the diagram the objects are indicated according to dataset and available photometry, together with evolutionary tracks and isochrones. In the next section we will analyze the accuracy of our estimate of the luminosity to quantify how much of the observed scatter can be due to observational uncertainties. Then, in Sect. \ref{distribution}, we will explicitly derive the age of each star through interpolation over the tracks and the isochrones to study the age distribution of our sample and to quantify the amount of scatter in terms of age spread.

% =========================
% 4. SOURCES OF UNCERTAINTIES
% =========================

\section{Sources of uncertainties} \label{source_error}

As mentioned before, there are concerns regarding the reliability of derived ages, as well as masses, from the HR diagram of young clusters, due both to the discrepancies between different PMS evolutionary models and to the various uncertainties affecting observations. Whereas finding the optimal model goes beyond the scope of this work, we can  take a fresh look at the observational errors listed by \citet{Hartmann2001} and discuss how they may affect our results on the ONC. First of all we present the analytical estimate of the uncertainties on luminosity and age, whereas in Sect. \ref{MCsim} we describe a Monte Carlo approach for the estimate of the same quantities.

\subsection{Analytical estimates}
\textit{Distance}. The VLBI (Very Long Baseline Array) distance obtained by measuring the trigonometric parallax of  radio sources in the Orion Nebula Cluster gives $d=414 \pm 7$~pc. The
standard deviation is larger than the transversal size (about 3~pc) of the cluster, and therefore still dominates over the intrinsic spread of distances within the cluster. We can therefore assume $\sigma(\log L)=0.015$.

\textit{Spectral type}. The spectral types given by \citet{Hillenbrand1997} are accurate to within one sub-type. According to \cite{Hartmann2001} errors on the spectral type tend to cancel; e.g a star which is erroneously assigned an earlier spectral type would have a larger age at a constant luminosity, but the extinction would be also overestimated with consequent erroneously increased luminosity. Averaging over the spectral type range of  K6-M1, \cite{Hartmann2001} found that an error within a sub-type in spectral type corresponds to an error on age of $\sigma(\log t)=0.02$. \cite{DaRio2010} has shown that even when varying the effective temperature within its error and consequently changing the bolometric correction in order to calculate the luminosity, due to the variation in extinction, the error on the luminosity is still parallel to the isochrone (see e.g. Figure 17 in \cite{DaRio2010}). 

\textit{Extinction}. \textit{Chorizos} provides a typical uncertainty on the extinction ($\sigma(A_{V})\sim$ 0.2 mag). Using equation~\ref{BC} to propagate errors, and knowing that $A_{I}\sim$0.5 $A_{V}$ and $A_{z}\sim$0.4 $A_{V}$, this translates into $\sigma(\log L)\sim$0.05.

\textit{Photometric Variability}. According to \citet{Herbst2002} the average variation in magnitude due to photometric variability for the ONC is $\Delta I=0.18$ mag. Following the same authors, we use the standard deviation to characterize the level of variability, since a general peak-to-peak measure can be affected by single accidental error. This standard deviation ($0.075$ mag) gives a logarithmic uncertainty on luminosity of $\sigma(\log L)= 0.03$.

\textit{Unresolved Binary companions}. With a pixel scale of 50 milliarcsec, ACS is able to resolve sources of comparable brightness with separation larger than $\sim50$~AU at the distance of the cluster. Assuming the standard lognormal period/separation distribution that peaks at approximately 30~AU \citep{Duquennoy1991}, and a binary fraction similar to the one observed in the field \citep{Reipurth2007}, we have a probability of 20\% of unresolved binary companions. With this probability the basic model described in \citet{Hartmann2001} allows us to derive a logarithmic error on luminosity due to unresolved binaries of $\sigma(\log L)=0.016$.

\textit{Accretion}. Our analysis allows us to derive the amount of accretion luminosity that affects each source and to subtract it. However, the error on $L_{acc}$ is given by \textit{Chorizos} in terms of $\sigma(\log(L_{accr}/L_{tot}))$, and consequently depends on $\sigma(\log L_{tot})$ which is what we want to determine. The best way to estimate the effect of this uncertainty on the luminosity is through Monte Carlo simulations. Therefore, we neglect this contribution in this preliminary calculation.

The quadratic sum of all these errors gives an overall uncertainty  $\sigma(\log L)=0.06$, which is roughly 1/3 of the value  $\sigma(\log L)=0.16$ reported by \citet{Hartmann2001} for the Taurus cloud. This reflects the better quality of our data set.

It is known that the ages of low-mass pre-main-sequence stars on the Hayashi tracks exhibit an approximate
power-law relation to the stellar luminosity L for each
mass, of the form $t\propto L^{-3/2}$ \citep{Hartmann2009}. Because of this relationship
between luminosity and age, the error
distribution in age takes the form $\delta t/t\sim\delta \log t$, with
\begin{equation}\label{logt}
\sigma (\log t)= (3/2) \sigma (\log L).
\end{equation}
Therefore, from the uncertainty on the luminosity one can  derive the error on the age, which turns out to be  $\sigma(\log t)=0.09$. As shown also by \citet{Hartmann2001},
errors of this order may preserve a time dependence of the star formation rate possibly present in the observations, a dependence which instead is lost when the observational errors are  $\sigma(\log t) \gtrsim $0.2.

Since our preliminary estimate of the overall uncertainty turns out to be smaller than the limit indicated by Hartmann, we can be fairly confident that, from the observational point of view, our uncertainties may allow us to infer real age trends from the analysis of the HR diagram. However, to understand to which extent this can be done, a more complete analysis is needed (see Sect. \ref{age_spread}).

\subsection{Monte Carlo simulations.}\label{MCsim}
After having determined a rough estimate of the uncertainty on luminosity, to better quantify the error on our age estimates, we have run Monte Carlo simulations to evaluate the dispersion in the estimated absolute luminosity of each object due to all the sources of uncertainty illustrated in the previous section. By generating artificial errors for each star we can more accurately determine the overall uncertainties on the the bolometric luminosity.

To evaluate the extinction and the accretion luminosity we have run \Cho varying the photometry according to the photometric errors and adding variability, following the variability distribution in the $I$ band given by \citet{Herbst2002}. We computed  the amount of variation in the other bands through synthetic photometry on artificial stars with standard spot temperatures, radii and filling factors for T Tauri stars \citep{Berdyugina2005}. Each time we have also varied the value of the distance within its error and added a shift in luminosity due to binarity, assuming the probability of unresolved companions given in Sect. \ref{source_error}.

We ran N=200 trials for each source, determining each time the error in the estimate of the luminosity and averaged the dispersion over the whole sample, obtaining values for $\sigma (\log L)$. We assume the value of $\sigma (\log L)=0.10$ as representative of the mean uncertainty we have on the luminosity.

According again to Equation~\ref{logt}, this error translates in an uncertainty on the estimate of the ages of  $\sigma (\log t)= 0.15$, which is larger than what we derived from basic error propagation  ($\sigma(\log t)=0.09$). A larger error is expected, since we are considering also the uncertainty in the estimate of the accretion luminosity, that was not kept into account in the previous calculation.  Still, this value is below the  $\sigma(\log t)=0.2$ threshold given by \citet{Hartmann2001}, and clearly remains much smaller than the observed scatter (see Sect. \ref{distribution}).

% =========================
% 5. THE AGE DISTRIBUTION
% =========================

\section{The age distribution} \label{distribution}
\begin{figure}
\includegraphics[scale=.33]{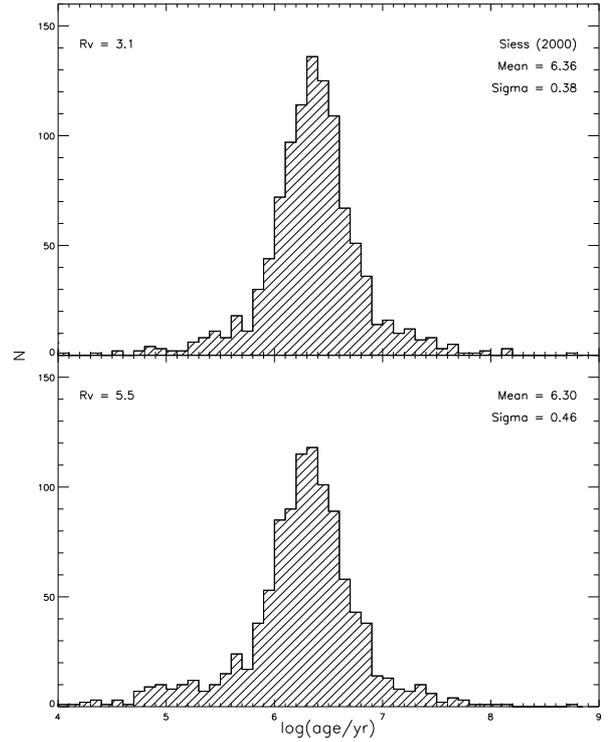}
\caption{\small{\textit{Distributions of stellar ages.} Histogram of the number of stars respect to the stellar ages, as derived from \cite{Siess2000} evolutionary calculations for the reddening laws $R_{V}=3.1$ (top) and $R_{V}=5.5$ (bottom).}}\label{age_distr_3.1}
\end{figure}

In  Fig. \ref{age_distr_3.1} we show the logarithmic distribution of stellar ages obtained for the two reddening laws we adopted and the \citet{Siess2000} evolutionary models. The average ages and age dispersions for the ONC that we derived with this set of evolutionary tracks and for the different reddening laws are listed in Table \ref{tab:age_mod}.

The cluster age turns out to depend on the reddening law. In general, ages derived assuming a reddening law of $R_{V}=3.1$ are larger than those derived assuming $R_{V}=5.5$. This is expected, since high values of $R_{V}$ imply higher extinction values and therefore higher intrinsic stellar luminosity and younger age. The mean age predicted by \citet{Siess2000} for the sample is $\sim$ $2.3$~Myr and $\sim$ $2$~Myr for $R_{V}=3.1$ and $R_{V}=5.5$, respectively.

\begin{table*} 
\caption{Mean age and age dispersion} 
\label{tab:age_mod}	
\centering \begin{tabular}{c c c c c} 
\hline\hline
   &   \multicolumn{2}{c}{$R_{V}=3.1$} &   \multicolumn{2}{c}{$R_{V}=5.5$}\\
\hline
\emph{PMS model}  & $<\log (age)>$ & $\sigma\log (age)$ & $<\log (age)>$ & $\sigma\log (age)$\\
\hline
Siess et al. (2000) &  6.36  &     0.38  &    6.30  &     0.46\\
D'Antona \& Mazzitelli (1998)&  5.93  &     0.59  &   5.82  &     0.71 \\
Palla \& Stahler (1999) &  6.14  &     0.49  &    6.07  &     0.53 \\
\hline
\end{tabular} 
\end{table*}

\begin{figure}
\includegraphics[scale=.33]{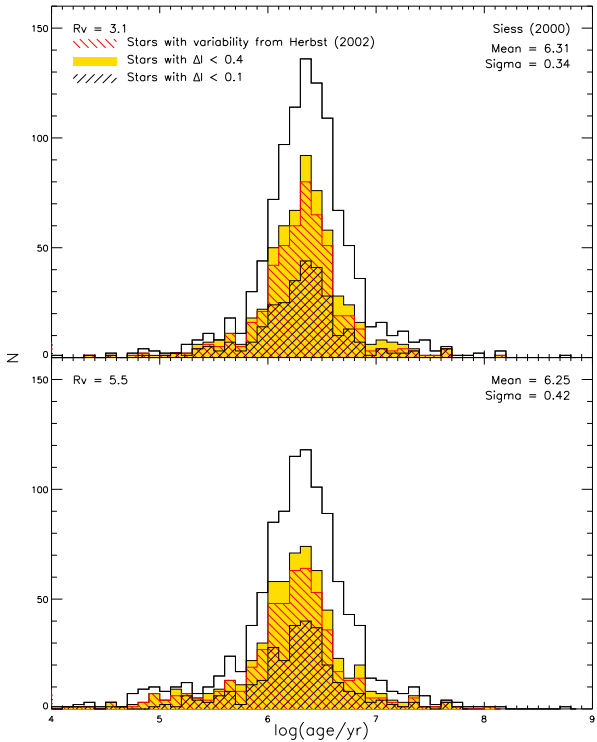}
\caption{\small{\textit{Distributions of stellar ages for low variability stars.} The stellar ages are derived from \citet{Siess2000} evolutionary calculations for the reddening laws $R_{V}=3.1$ (top) and $R_{V}=5.5$ (bottom). The open histogram represents the total sample. The yellow area indicates the sample of 608 stars for which we have information about variability from \citet{Herbst2002}. The red hatched histogram represents the sources selected with $\Delta I\leq 0.4$ mag while the black one the stars with $\Delta I\leq 0.1$. The average age of the cluster and the age spread are relative to the third sample.}}\label{age_sel_distr}
\end{figure}

Concerning variability, using the information supplied by \citet{Herbst2002}, it is possible to select a sub-sample of 508 objects showing a peak-to-peak variability $\Delta I \leq 0.4 $ and 298 objects with $\Delta I\leq 0.1$. For these two samples the total error in the logarithm of the luminosity, as estimated through basic error propagation, goes to $\sigma(\log L)=0.053$ and $\sigma(\log L)=0.051$, respectively (for the complete sample $\sigma(\log L)=0.06$, see Sect. \ref{source_error}). 
In Fig. \ref{age_sel_distr} we show the distributions of the ages for the different samples and in Table \ref{tab:age_sel} we summarize the results. A direct comparison with the estimates obtained for the full sample does not show substantial differences both in the average age and age spread.
The Kolmogorov-Smirnov test (KS test), which returns the probability that two distributions were drawn from the same parent sample, confirms that the samples are consistent with the original population, ie. variability does not affect significantly the mean and the spread of the age distribution.

\begin{figure}
\includegraphics[scale=0.33]{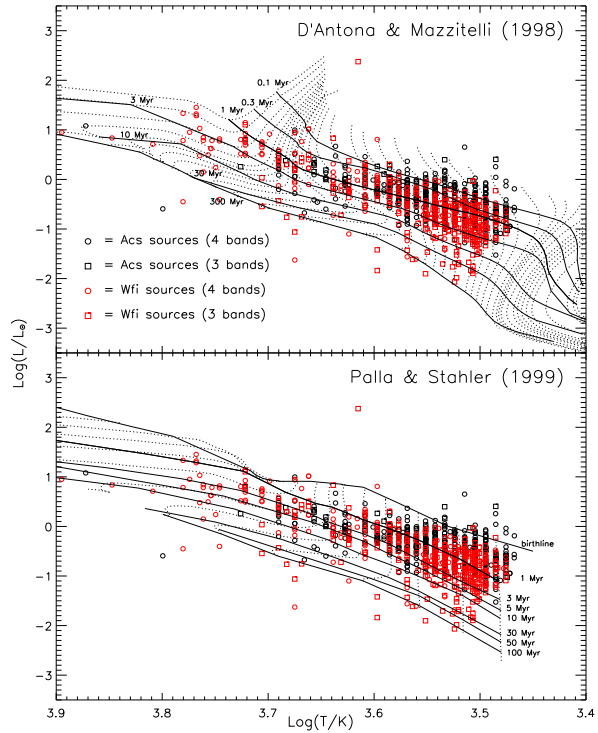}
\caption{\small{\textit{HR diagrams for the Orion Nebula Cluster}. Same legend as in Fig. \ref{hr_diagr_siess}. From the top to the bottom are superimposed evolutionary tracks and isochrone from \cite{Dantona1997} and \cite{Palla1999}. Dotted lines represent evolutionary tracks and they range from 0.017 to 3 $M_{\odot}$ for D'Antona \& Mazzitelli and from 0.1 to 6 $M_{\odot}$ for Palla \& Stahler.}}\label{hr_diagr_tot}
\end{figure}

\begin{figure}
\includegraphics[scale=.33]{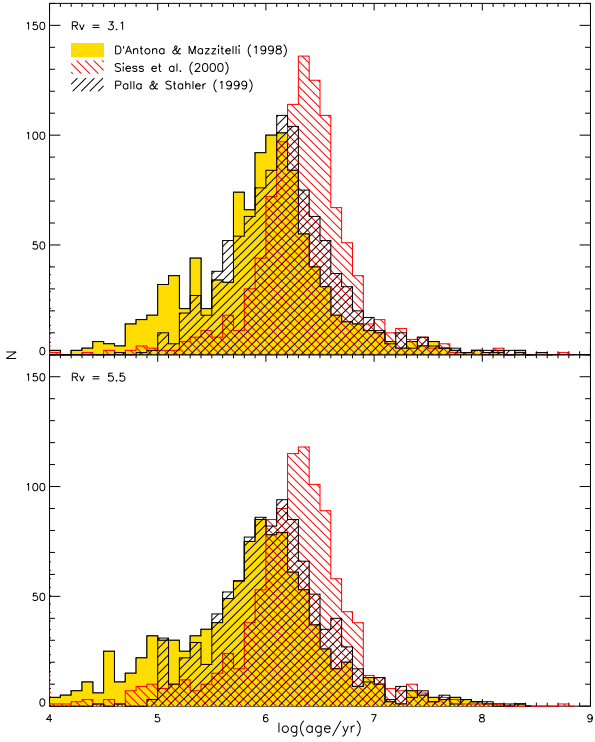}
\caption{\small{\textit{Distributions of stellar ages for different PMS evolutionary models.} These histograms are relative to the reddening laws $R_{V}=3.1$ (top) and $R_{V}=5.5$ (bottom). The yellow histogram represents the results obtained with interpolation on \citet{Dantona1997} isochrones. The red hatched histogram is relative to \citet{Siess2000} while the black hatched one shows \citet{Palla1999} predictions.}}\label{age_distr_mod}
\end{figure}

As mentioned before, we have considered three different families of PMS evolutionary tracks and isochrones: \citet{Dantona1997}, \citet{Siess2000} and \citet{Palla1999}. In Figs \ref{hr_diagr_tot} and \ref{age_distr_mod} we present the HR diagram and the age distribution according to these different sets and in Table \ref{tab:age_mod} the mean ages and the age spreads predicted for the Orion Nebula Cluster. The differences in the estimate are clearly visible. First of all, this discrepancy affects both the mean age of the cluster and the age of the single star. Regardless of the value of the reddening parameter, \citet{Dantona1997} consistently predicts younger ages, which appear to be consistent with the results obtained for the same evolutionary calculations by \cite{Hillenbrand1997}. Second, the value of the mean age depends also on the predictive capability of the model for this particular cluster. Since the different sets cover slightly different areas in the HR diagram (Fig. \ref{hr_diagr_tot}), certain stars may or may not be left out of certain models. Finally, a relative age spread appears to be robust regardless of which family of models is used, although the amount of this spread is model dependent. The scatter predicted by \citet{Siess2000} is the smallest one with both the standard and the anomalous reddening law.

\begin{table*} 
\caption{Selected samples: mean age and age dispersion (Siess et al. 2000)} 
\label{tab:age_sel}
\centering \begin{tabular}{c c c c c c c c c c} 
\hline\hline
   &  \multicolumn{2}{c}{Herbst Sample} &   &  \multicolumn{2}{c}{First selection} &  & \multicolumn{2}{c}{Second selection} \\
\cline{2-3} \cline{5-6} \cline{8-9} 
\emph{$R_{V}$}  & $<\log (age)>$ & $\sigma\log (age)$ &  & $<\log (age)>$ & $\sigma\log (age)$&   &$<\log (age)>$ & $\sigma\log (age)$\\
\hline
3.1 &  6.33  &     0.36  &   &   6.31  &     0.34  &     & 6.31  &     0.36\\
5.5 & 6.37 &     0.43  &   &      6.25  &     0.42  &     & 6.27  &     0.41 \\
\hline
\end{tabular} 
\end{table*}

In conclusion, whereas from the observational point of view our sample enables us to place the stars in the HR diagram with fairly high precision, different PMS evolutionary models provide significantly different ages. But regardless of the choice of the set of models, there appears to be an age scatter which cannot be accounted for by observational errors. Star formation in the ONC has mainly occurred within the last 10 Myr, with $\sim$ 70\% of the stars younger than 3 Myr. There is no evidence of a single star formation burst as the distribution spans a wide range of ages with a scatter of a few Myrs, in agreement with what was found by \citet{Hillenbrand1997} and \citet{DaRio2010}.

\subsection{Completeness Analysis}\label{completeness}
The HRD and the age distribution presented here, are relative to the subsample of ONC objects with at least three-band photometry and known spectral type. This represents only a fraction of our entire photometric catalog of the ONC, and its completeness, which worsen towards fainter luminosities -- or lower masses -- is basically depending on the availability of spectral types, and not on the photometric detection. This is because, whereas our HST photometry extends well into the BD mass range and goes 5 magnitudes fainter than \cite{Hillenbrand1997}, spectral types reach at most the lowest masses in the stellar regime.

Towards this limit, the subsample becomes also extinction limited, in the sense that very-low mass members will have an available $T_{\rm eff}$ estimate only if their $A_V$ is not significantly large. The reason for this comes from source of our spectral types: the majority of these are from the optical spectroscopy of \citet{Hillenbrand1997}; in that work optical spectra were obtained selecting randomly the candidates, but only for sources with previously available $V$ and $I$ photometry, and clearly were limited to lower magnitude limits. Other spectral types were derived by \citet{DaRio2009} 
using a photometric index computed from ground-based measures in the $V$ and $I$ bands, plus a medium-band filter centered on the TiO feature at $\lambda\sim6200\AA$; this second set reaches spectral types as late as M6, for moderate extinction.

Besides the optical luminosity constraint, no additional criterion was imposed (both in \citet{Hillenbrand1997} and \citet{DaRio2009}) for the selection of sources with assigned spectral types. Consequently, the only selection effect which affects the completeness of our subsample is the apparent $V$ magnitude of the members, since due to the
intrinsic color and reddening of typical ONC populations the detection is  generally easier at longer wavelengths.

Therefore, we computed the completeness distribution of our subsample as a function of $V$ magnitude. Even at the faintest end of range of $V$ magnitudes, the ACS photometric catalog can be considered complete reaching several magnitudes deeper. Therefore, the ratio of the two $V$-band luminosity functions (in the subsample and for the whole ACS catalog) provides automatically the $V$-band completeness function. More precisely, this computation was done separately in three subcases: the ACS stars, WFI stars in the FOV of ACS, and the WFI stars outside the ACS FOV. Since at the bright end of the LF, the ACS catalog is deficient of sources with respect to WFI, due to saturation, we complemented the ACS catalog with the missing bright WFI sources falling in the same area. By adding WFI sources outside the ACS FOV, we are allowing the completeness function to be $>100\%$ for some bright $V$-magnitudes; however, this is not a problem, since any multiplicative re-normalization of the completeness function is absolutely arbitrary and irrelevant for our purposes.

We want now to pass from the $V$-band completeness to the completeness as a function of stellar parameters, like $T_{\rm eff}$ and $\log L$, or better, mass and age. This cannot be done directly, because differential reddening, as we said, introduces an extinction limited bias. The computation is still possible in a probabilistic way, through a monte-carlo simulation. The method is identical to that used in \citet{DaRio2010}: we consider a pair of values of mass and age. Assuming the evolutionary models, we convert this point in $T_{\rm eff}$ and $\log L$. Then we apply backwards bolometric corrections and intrinsic colors to covert this point further to the $V$ vs $V-I$ CMD, assuming $A_V=0$. Having the intrinsic $V_0$ magnitude of the $(mass,age)$ point, we redden this test star randomly, with values of $A_V$ drawn from the reddening distribution of the ONC, computed on the brightest half of the population. Each of this ``reddened'' test stars will have a different (fainter) $V$ magnitude than $V_0$; and therefore a different completeness. By simply averaging all these completeness ratios, we derive the exact completeness of the point in the mass-age plane considered at first. Iterating this procedure for each (mass,age) pair, we derive a 2-dimensional completeness function in the mass-age plane.

\begin{figure}
\includegraphics[scale=0.33]{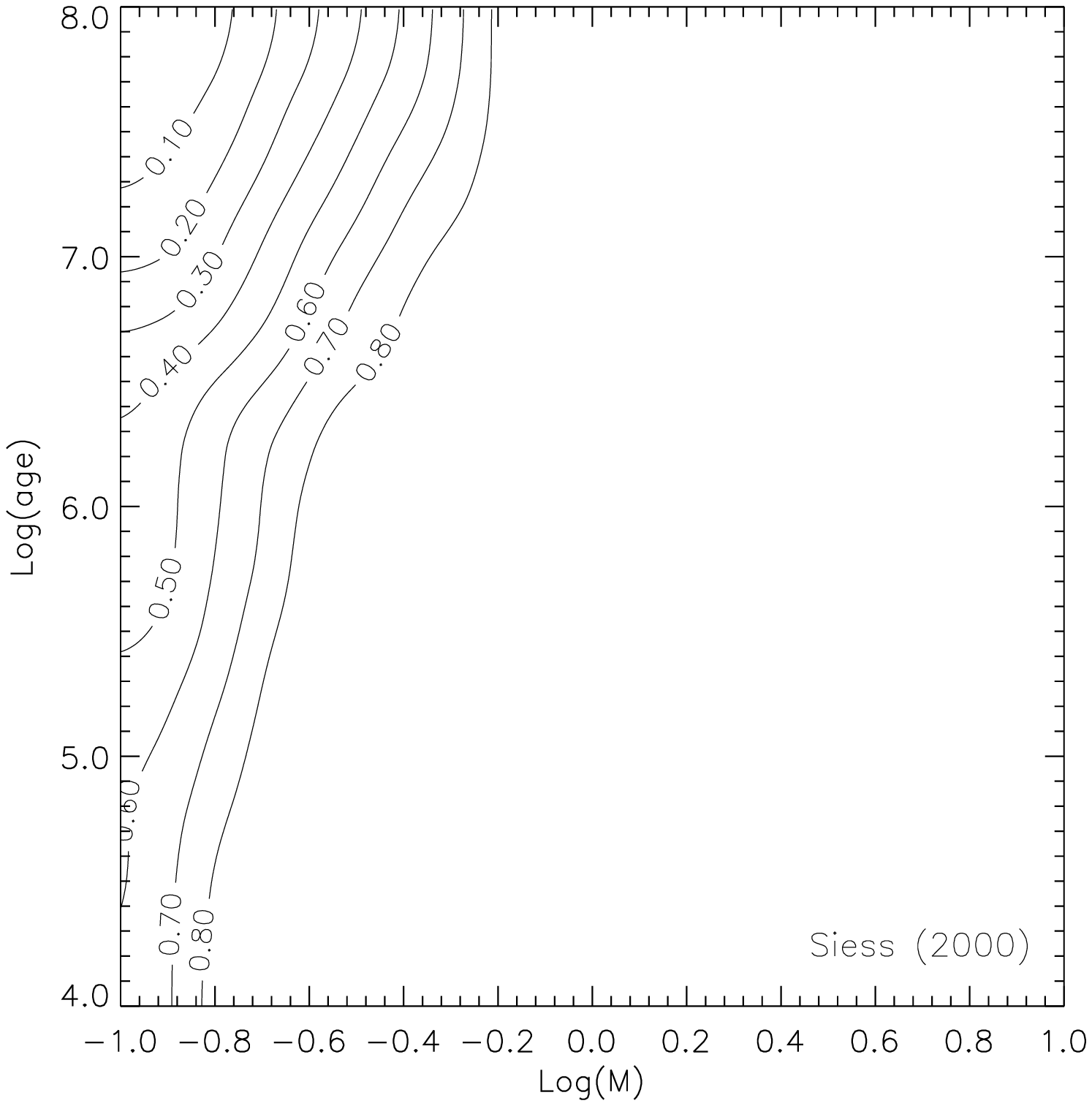}
\includegraphics[scale=0.33]{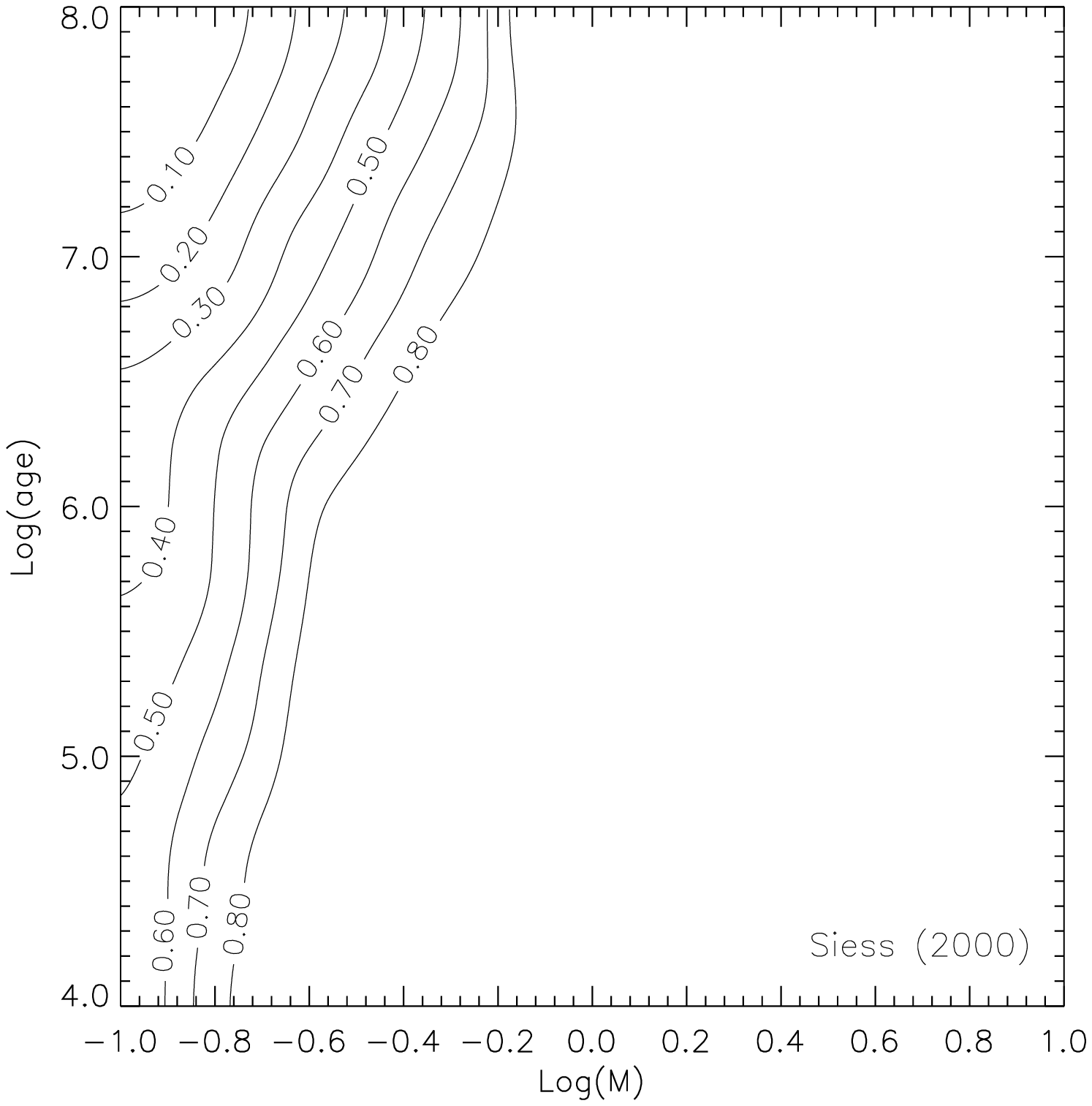}
\caption{\small{Contour plots of the completeness function in the mass/age plane, computed from our simulations for the evolutionary models of \cite{Siess2000} and for $R_{V}$ =3.1 (\textit{top}) and $R_{V}$ =5.5 (\textit{bottom}). }}\label{compl_contour}
\end{figure}

\begin{figure}
\includegraphics[scale=0.38]{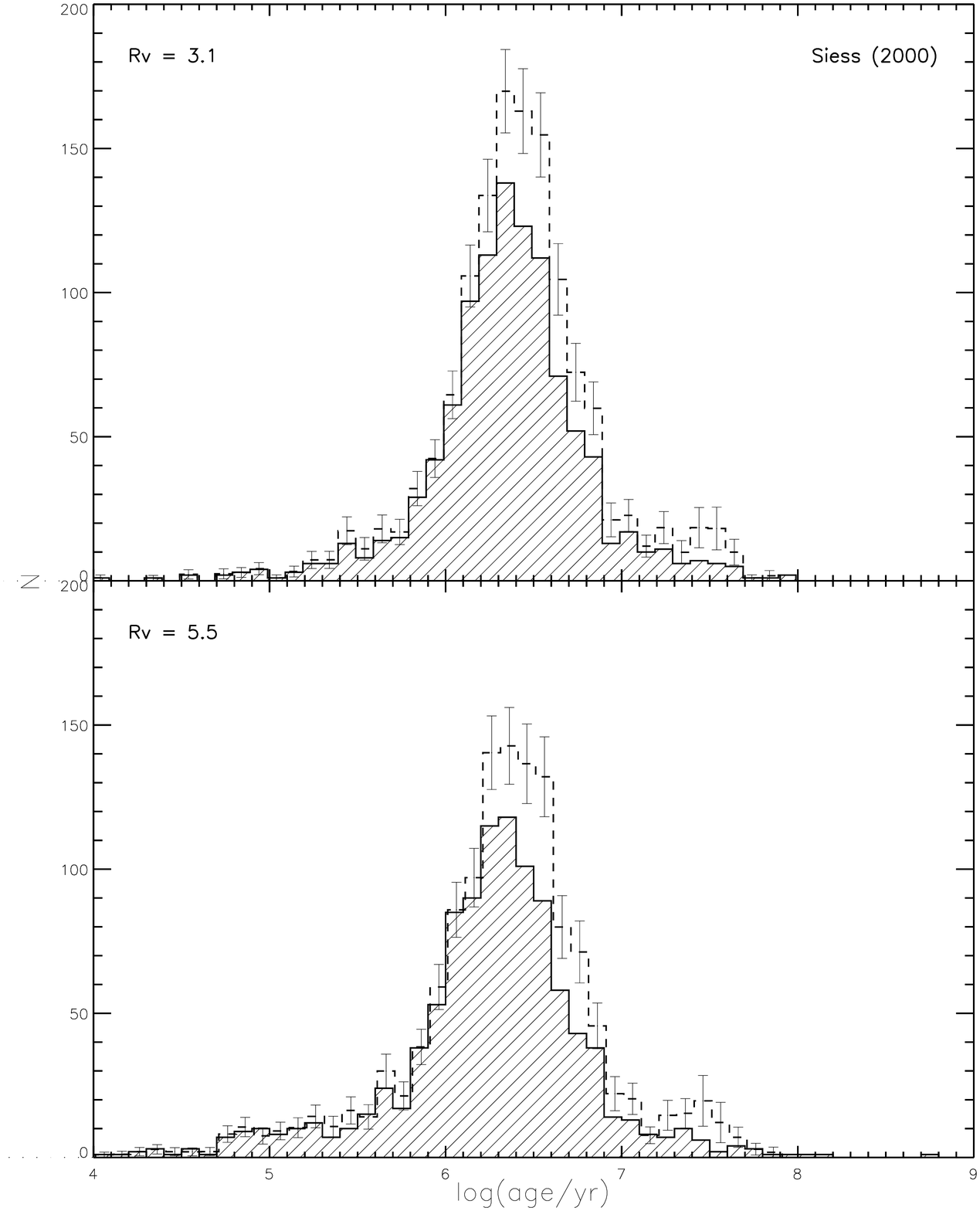}
\caption{\small{\textit{Completeness-corrected age distribution}. The hatched histogram represents the age distribution of our sample. The open histogram shows the age distribution of the completeness-corrected sample}}\label{compl_hist}
\end{figure}

Figure \ref{compl_contour} shows the completeness contours in the mass-age plane for the two different reddening laws, whereas in Figure \ref{compl_hist} we present the completeness-corrected age distributions according to \cite{Siess2000} evolutionary models for the two different reddening laws. We observe that completeness decreases toward lower masses and higher ages, therefore the completeness correction does not change significantly the average ages and the measured age spread ($<\log (age)>$=6.40 and $\sigma\log (age)$=0.49 for $R_{V}$=3.1) but increases the old age end of the distributions.

However, optical photometry alone is often not enough to identify the complete stellar population in young clusters. Near-infrared surveys of the ONC in the past decades have revealed the presence of a large number of highly extincted sources in this region \citep[e.g.][]{Hillenbrand1998, Muench2002}, even though a more recent near-infrared study has shown that a majority of stars ($\sim$70\%) of the ONC has $A_V<$5 (\cite{Robberto2010}, Scandariato et al. in prep.). Moreover the contamination becomes an important issue with increasing sensitivity of the surveys. Therefore it is not correct to assume that everything that has been observed in the near-infrared is a cluster member.
In any case, because our survey is not \textit{complete} down to 0.25 $M_{\odot}$ for more deeply embedded association members, we have tested whether our sample is biased towards older stars.
We considered a mass-age-$A_V$ limited sample (M$<$0.25$M_{\odot}$, $t<10^7$ Myrs and $A_V<$5) of 647 stars in total. We divided this subsample in two halves with equal number of objects according to low ($A_V\lesssim$2) and high ($A_V\gtrsim$2) extinction. We have used the KS test to check whether the age distributions of the two samples are drawn from the same parent population and we are unable to reject the hypothesis that the distribution are the same ($P(d)_{KS}$=0.76).
Since we do not see evidence for a trend of age with increasing $A_V$, we argue that having most of the stars, though not all, in our sample allows us to study the age distribution of the Cluster based on a lower extinction sub-sample.

\subsection{Membership}
It could be argued that some of the scatter observed in the HR diagram and the age distribution might be due to non-member contamination.
The Orion complex itself consists of three subgroups, the ONC (Orion subgroup Id) and other three subgroups of stars (Ia, Ib, and Ic, \cite{Brown1994}). The most likely subgroup from which we would see contamination in our data is the Orion subgroup Ic, since it is located along the same line of sight of the ONC. Orion Ic is thought to be less than 5 Myr old according to \cite{Brown1994}, thus we would expect little differences between the isochrones of the two subgroups.

\begin{figure}
\includegraphics[scale=0.38]{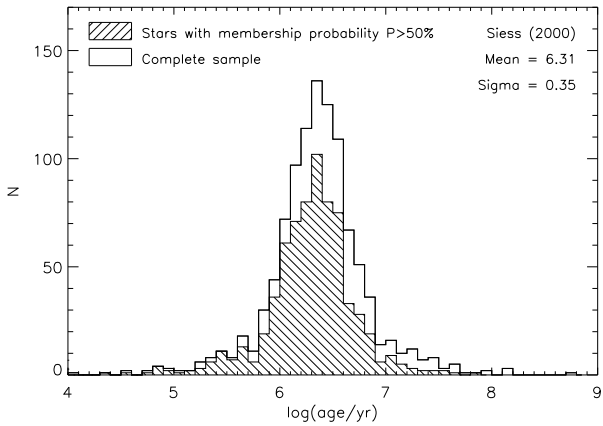}
\caption{\small{\textit{Age distribution of confirmed members}. The open histogram represents the age distribution of the complete sample. The hatched histogram shows the age distribution for the  stars also present in the H97 catalog with membership probability P$>$50\% according to the proper motion survey of \cite{Jones1988}. The KS test gives a probability of 17\%  that the two distributions are consistent. }}\label{member_distr}
\end{figure}

In any case,  for the ONC objects also present in H97 we utilized the membership given in that work, which was taken from the proper motion information of \cite{Jones1988}, to study the age distribution of the confirmed members.  Among our 1057 stars, 701 have proper motion measurements and are thought to be ONC members with a probability P $>$50\%. 
In Fig. \ref{member_distr} we show the age distribution of the whole sample (open histogram) and of the 703 confirmed members (hatched histogram) in the case of $R_{V}$=3.1 and \cite{Siess2000} evolutionary models. The KS test give a probability of 17\% that the two distribution are drawn from the same parent population and the spread ($\sigma\log (age)$=0.40) obtained for the subsample does not differ significantly from $\sigma\log (age)$=0.43 derived for the whole sample.

For these reasons we are confident that contamination does not contribute very much to the scatter observed in the age distribution of the ONC.

% =========================
% 6. AGE SPREAD
% =========================

\section{Age spread} \label{age_spread}
The observed dispersion in age represents an upper limit to the real age spread, being a combination of true age spread and scatter due to observational uncertainties. These, however, cannot explain the scatter of $\sim 1$ dex observed in the HR diagram, as we have found in the previous sections. To quantify, or at least to put a constraint, on the real age spread it is necessary to run a second Monte Carlo simulation accounting for observational errors. 

\begin{figure}
\includegraphics[scale=0.38]{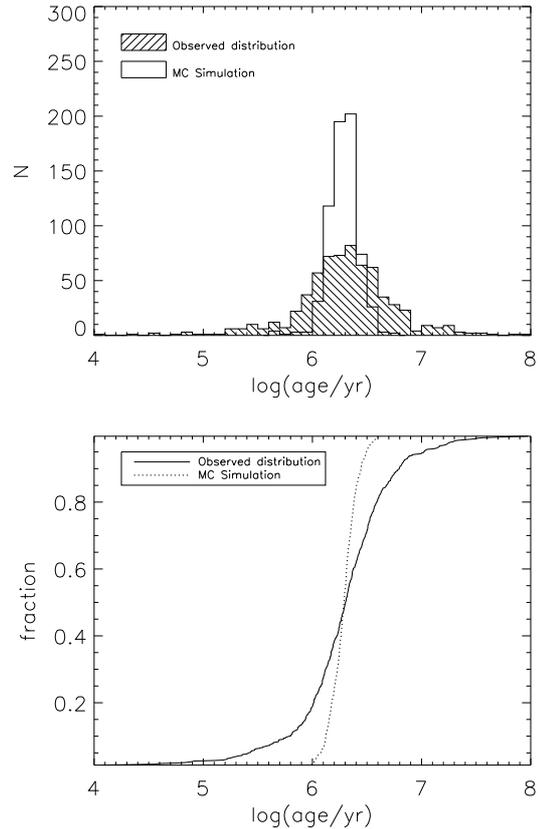}
\caption{\small{\textit{Test of coeval population}. (\textit{Top}) The open histogram represents the age distribution simulated for a sample stars uniformly distributed along a 2.2 Myr isochrone from \citet{Siess2000}, assuming an observational uncertainty on the luminosity of $\sigma (\log L)=0.10$. The spread in age is $\sigma (\log t)=0.13$. The hatched histogram shows the observed age distribution for the observed completeness-limited subsample. (\textit{Bottom}) Cumulative functions of the the two distributions, solid and dotted line represent the observed and simulated sample respectively.}}\label{test_isochrone}
\end{figure}

In order to increase the reliability of our results, we limited the analysis in the mass range in which our completeness is above 80\%.
Our first step has been to test the hypothesis of an exactly coeval cluster, simulating a sample of stars with luminosities and temperature corresponding to a 2.2 Myr isochrone (mean age value for the ONC estimated in Sect. \ref{distribution}) according to \cite{Siess2000} evolutionary model. We have then varied their luminosities assuming a Gaussian distribution of errors with $\sigma (\log L)=0.10$ (see Sect. \ref{source_error}), recovered their ages from interpolation over the \cite{Siess2000} isochrones and evolutionary tracks and studied the overall age distribution. The spread around the isochrone is $\sigma (\log t)=0.13$. We compared this distribution with the observed one using the KS test and we found a probability of $10^{-19}$ that the two distributions are drawn from the same parent population. The comparison is shown in Fig. \ref{test_isochrone}. This formally confirms that the observational uncertainties alone are not able to account for the age spread observed in the region.

\begin{figure*}
\resizebox{\hsize}{!}{\includegraphics[scale=0.38]{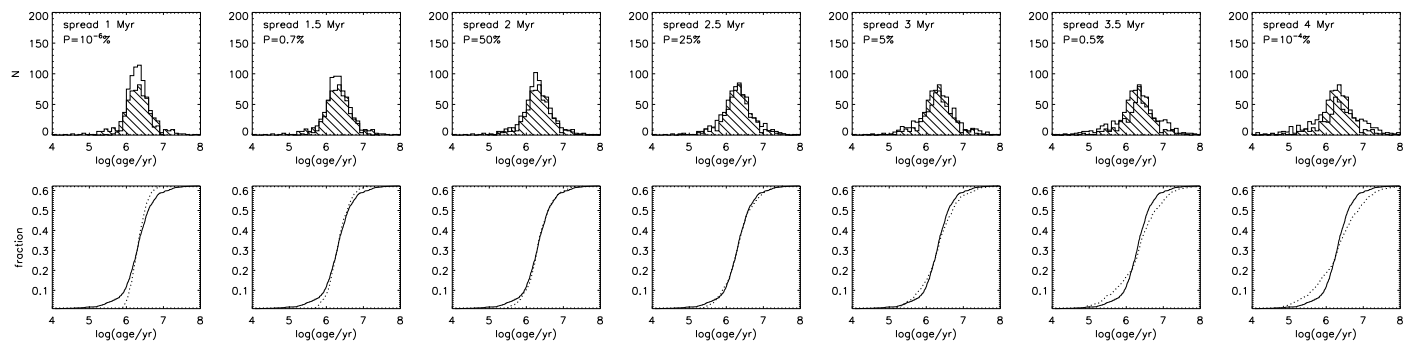}}
\caption{\small{\textit{Age spread test}.  (\textit{Top, from left to right}) The open histograms represent the age distribution simulated for a sample stars with mean age of 2.2 Myr, assuming an observational uncertainty on the luminosity of $\sigma (\log L)=0.10$ and with intrinsic age spreads of 1, 1.5, 2, 1.5, 3, 1.5 and 4 Myr respectively. The KS test probability between the two distributions is also given. (\textit{Bottom}) Cumulative functions for each simulation. We use the same legend as in Fig \ref{test_isochrone}.}}\label{test_age_spread}
\end{figure*}

We then performed other simulations to find the amount of age spread needed to reproduce the observations, in addition to the dispersion due to observational uncertainties. In Fig. \ref{test_age_spread} we show the comparison with the simulations in the cases of spreads of 1, 1.5, 2, 2.5, 3, 3.5 and 4 Myrs. The probabilities given by the KS test are also shown in each panel of Fig. \ref{test_age_spread}. Since for spread larger than 3.5 and smaller than 1.5 Myr the KS test probability drops below 1\%, we suggest the real age spread to be within 1.5 and 3.5 Myr. 

Therefore, on the basis of the isochrones, we find a spread of roughly 2-3 Myr in age, which is in good agreement with the results presented by \cite{DaRio2010}, and other techniques for inferring stellar ages also confirm the same results \citep{Muench2008}. 
Another way to interpret this result is that we observe a scatter in stellar radii at any given mass. The same evidence was found by \citet{Jeffries2007} using the rotation periods and projected equatorial velocities ($v$ $\sin i$) for a sample of stars in the ONC.  \citet{Jeffries2007} studied the distribution of $R$ $\sin i$ and modeled it as dispersion of true radii, inferring a spread in stellar ages. This spread is again approximately a few Myrs. 

With reference to the recent attempts, \citet{Baraffe2009}, \citet{Jeffries2011} and \citet{Littlefair2011} with different methods suggest that the spread in radius observed in different star forming regions may be due to the accretion history and not to a real spread in age. \citet{Baraffe2009} present a theoretical calculation for the expected scatter in luminosity due to episodic accretion, while \citet{Jeffries2011} find that stars with different ages do not show differences in accretion and disk excess and propose they have the same age. Finally, \citet{Littlefair2011} observe a correlation between rotation velocity and luminosity, which is opposite in respect to what is expected in the case of a real age spread. This, according to the authors, might be supportive of the scenario presented in \citet{Baraffe2009}.
\cite{Hosokawa2011}, on the other hand, show that the broadening of the PMS sequence in the HRD due to episodic accretion, as computed by \citet{Baraffe2009}, was significantly overestimated, indicating that even this scenario might be incomplete. 

At the moment we do not exclude a superposition of a real age spread and episodic accretion, especially since a model quantifying whether the accretion-induced luminosity spread is able alone to account for the observations has not been developed yet. In any case a more complete scenario still needs to be investigated.

We remark, finally, that the Orion Nebula hosts a few sites of ongoing star formation, like e.g. the BN/KL system, Ori-1 South, and an extremely dense filament of gas and dust stretching through the center of the ONC which hosts at least 45 protostellar candidates \citep{Muench2008}.
In our study of the optically visible population we could not taken into account these embedded regions.
Our finding that the mean age of the ONC is 2 Myr and that the star formation appears to have occurred between $\sim$1-3 Myr ago is consistent with the currently detected level of ongoing star formation. However, if future observations reveal dozens more deeply embedded protostars over the same field of view as we observe several hundreds of T Tauri stars with ages of 1-3 Myr, the overall picture might change.

% =========================
% 7. SPATIAL AGE DISTRIBUTION
% =========================

\section{Spatial age distribution} \label{spatial_age_distribution}
In Sect. \ref{distribution} we derived the age distribution for the ONC according to different evolutionary models and reddening laws.  
We discuss now whether this distribution holds true at different location in the cluster.

\begin{figure*}
\resizebox{\hsize}{!}{\includegraphics[scale=0.3]{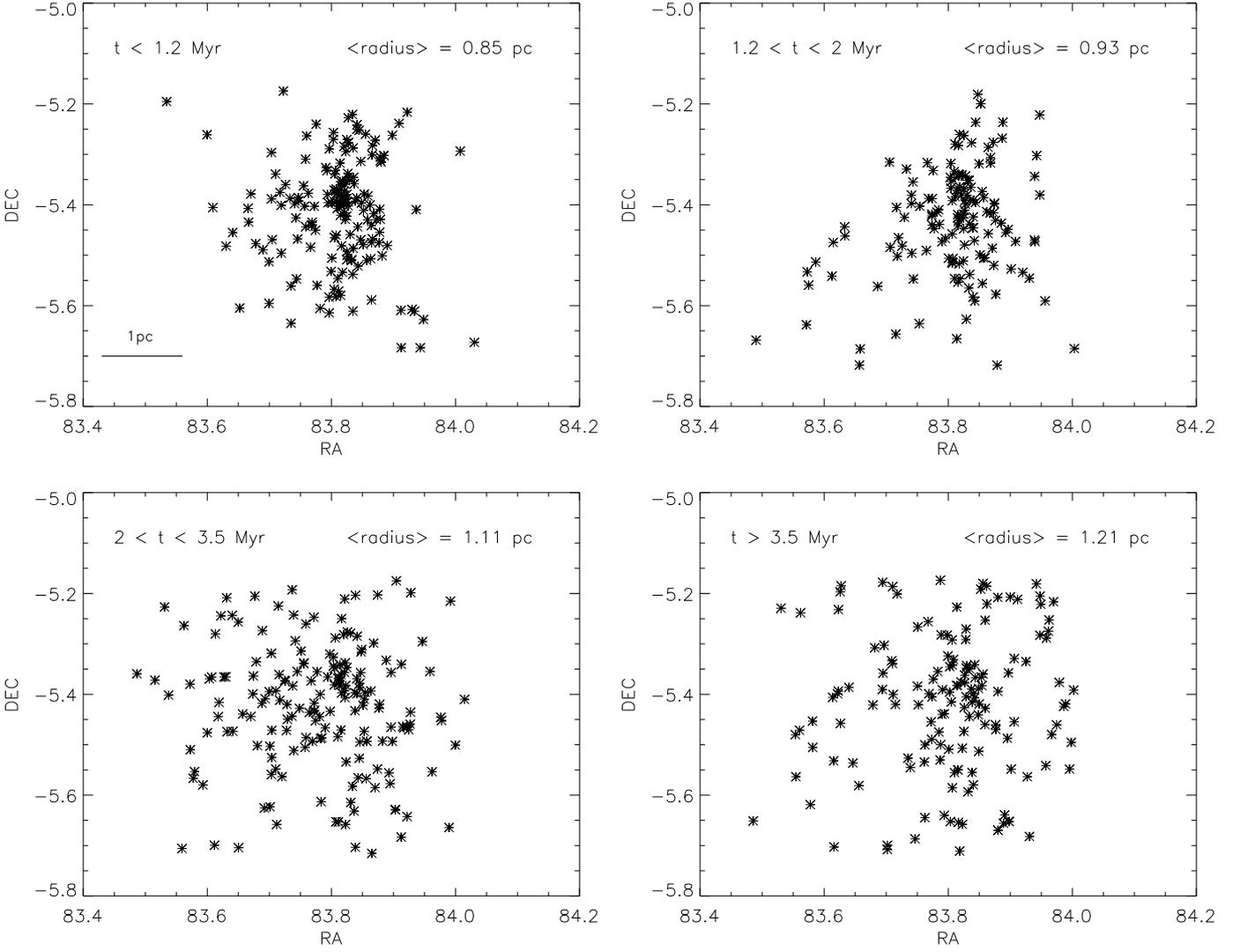}}
\caption{\small{\textit{Spatial age distribution}. The four images show the spatial distribution of stars within various age ranges. The youngest stars are clearly more centered towards the center than the oldest ones.}}\label{spc_age_distr}
\end{figure*}

We limit again our investigation of the cluster age structure to the subsample of stars for which our completeness is above 80\%. 
In Figure \ref{spc_age_distr} we show the spatial distribution on the sky of this subsample of ONC stars, binned by age. Each bin has roughly the same number of stars. As can be easily seen, the youngest stars are more clustered towards the center, while the oldest ones are almost homogeneously distributed in space. This evidence for age segregation was already noticed by \cite{Hillenbrand1997}.  A two dimensional KS test also confirms this first-look impression. According to the KS test, whereas comparing the spatial distributions of stars younger than the mean age of the cluster (first two panels) we get a probability of 16\% that they are consistent with each other, we find a probability $<$1\% that they are in agreement with the spatial distributions of the stars in the last two bins. Moreover, more than 65\% of the stars younger than 1 Myr are located within one parsec from the central O star, $\theta^{1}$ Ori C, commonly assumed as center of the cluster.

This preliminary result suggests a possible trend of ages
with spatial distribution. However, a part of this trend might be biased by spatially variable incompleteness.
In fact, while our completeness function  is representative,
on average, of the whole stellar sample we considered, the completeness varies with position. Specifically, it is worse towards the center of the system where the sky background is brighter.
In any case, if this ''age segregation'' is confirmed, this may have implications for the study of the
dynamical evolution of the cluster. 
 
\section{Dynamical evolution of the ONC} \label{dynamics}
In the past years there have been many attempts to understand the dynamical history of the ONC \citep[e.g.][]{Hillenbrand1998,Allison2009}.
In particular the evidence of mass segregation \citep{Hillenbrand1997,Hillenbrand1998} in such a young cluster highlights the problem of distinguishing between primordial and dynamical mass segregation.
In fact, the cluster crossing time is $\sim 2$ Myr, roughly the same mean age of the cluster, and an even longer time is required for the cluster to relax \citep{Hillenbrand1998}.
Thus, mass segregation could be primordial, meaning that the ONC formed with the most massive stars concentrated near the center. On the other hand, simulations of initially cool and clumpy clusters have shown that with these initial conditions mass segregation could also be dynamical on very short timescales \citep{Allison2009}. 
However if the observed spread in radii can be interpreted as a true age spread and the star formation in the ONC lasted for at least 2 Myr, the dynamics for the youngest and oldest stars could turn out to be very different:
the evolution of the oldest, earliest stars was dominated by the molecular cloud in which they were embedded, whereas the youngest stars in the clusters were subject to the dynamical influence of the stars already formed.This highlights that models which attempt to explain dynamical evolution in this region are probably still incomplete.

% =========================
% 8. CONCLUSIONS
% =========================
\section{Conclusions} \label{conclusion}
In this work, taking advantage of the broad-band photometry extracted from the large data-set of  ACS/WFC\ observations collected for the HST\ Treasury Program on the ONC, we have derived the HR diagram for the Cluster, with the main goal of studying the age distribution and age spread in the region.

Our method is based on the estimate of the extinction and accretion luminosity towards each source by performing synthetic photometry on an empirical calibration of atmospheric models \citep{DaRio2010} using the package \Cho \citep{Maiz2004}. This tool allows us to derive accurate estimate of the intrinsic luminosity for each star.
Taking into account uncertainties on the distance, spectral type, extinction, unresolved binaries, accretion and photometric variability we can determine luminosity and age with logarithmic errors of $\sigma(\log l)=0.10$ and $\sigma(\log t)=0.15$, respectively.  Compare to \citet{Hartmann2001}, we obtain errors small enough to preserve a time dependence of the star formation rate possibly present in the observations, which is instead lost when the observational errors are  $\sigma(\log t) \gtrsim $0.2.

The HR diagram we derived shows a large spread in luminosity. According to \cite{Siess2000} evolutionary calculations the mean age of the cluster is $2.2$ Myr, and it is slightly dependent on the assumed reddening law. Much stronger is the dependence on the evolutionary models since choice of different isochrones and tracks affects both the mean age of the Cluster and the estimate of the age of the single star. 
Currently there are several evolutionary models available and our capability of indicating the best one is limited to secondary clues (e.g. spacial distribution and kinematics vs. average cluster age) and our prejudice (e.g. lack of mass-age relation).
However, the presence of an observed spread larger than the errors on the age is common to all the models. 

A preliminary study of the spatial distribution of the ages suggests age segregation in the ONC. However we cannot exclude the possibility of an apparent trend, due to a spatial dependence of completeness. A further investigation is certainly needed. 

To quantify the amount of intrinsic luminosity spread in terms of age spread, we ran also MC simulations taking into account the sources of errors mentioned before. First of all, the observations are not consistent with a coeval stellar population, even if we account for the apparent age spread due to observational uncertainties. Second, the observed age distribution in the ONC is consistent with simulations of stellar population with intrinsic age spread between 1.5 and 3.5 Myrs, but not larger. 
These results confirm the presence of a real luminosity spread which cannot be accounted for only with observational uncertainties but reflects a real spread in the stellar radii for a given mass. We interpret this a true age spread in Orion, implying that star formation in this region lasted a couple of Myr. Some authors suggest that this spread might be due to the accretion history but at the moment the truthfulness of this scenario has not been proved yet. However any possible explanation for the spread observed in the HR diagram should reproduce this amount of age spread.
Within the current limitations, we conclude that the age is $2$ Myr and the age spread is $\sim 2$ Myr.

\begin{acknowledgements} 
We thank an anonymous referee for valuable suggestions.
Support for program 10246 was provided by NASA through a grant from the Space Telescope Science Institute (STScI), which is operated by the Association of Universities for Research in Astronomy, Inc., under NASA contract NAS 5-26555. Portion of the work of M. Reggiani was done under the auspices of the Summer Student Program of the Space Telescope Science Institute.
L. R. acknowledges the PhD fellowship of the International Max-Planck-Research School.
\end{acknowledgements}

% =========================
% THE BIBLIOGRAPHY
% =========================

\clearpage

\end{document}